\documentclass[pra,twocolumn,superscriptaddress,10pt,longbibliography]{revtex4-1}

\usepackage[utf8]{inputenc}
\usepackage[T1]{fontenc}
\usepackage{amsmath}    
\usepackage{graphicx}   
\usepackage{verbatim}   
\usepackage{color}      
\usepackage{subfigure}  
\usepackage{braket}
\usepackage[hidelinks]{hyperref}
\usepackage[capitalize]{cleveref}
\usepackage{booktabs}
\usepackage{MnSymbol}
\usepackage{siunitx}
\DeclareSIUnit\gauss{G}
\DeclareSIUnit\bohr{a_{B}}
\usepackage{bbm}
\usepackage{ulem}

\usepackage{color}
\definecolor{mygreen}{rgb}{0,0.5,0}
\definecolor{mygrey}{rgb}{0.5,0.5,0.5}
\definecolor{myred}{rgb}{0.75,0,0}
\definecolor{myblue}{rgb}{0,0,0.75}
\definecolor{mymagenta}{cmyk}{0,1,0,0.12}
\definecolor{mycyan}{cmyk}{1,0,0,0.12}
\definecolor{myorange}{rgb}{1,0.5,0}
\definecolor{myviolet}{rgb}{0.5,0.0,0.75}

\newcommand{\subeff}{_{\rm eff}}
\newcommand{\subint}{_{\rm int}}
\renewcommand{\subint}{_{\rm spin}}
\newcommand{\subhfs}{_{\rm hfs}}

\newcommand{\coll}{\smilefrown}

\newcommand{\ICFOAddress}{ICFO-Institut de Ci\`encies Fot\`oniques, The Barcelona Institute of Science and Technology, 08860 Castelldefels (Barcelona), Spain} 
\newcommand{\ICREAAddress}{ICREA -- Instituci\'{o} Catalana de Recerca i Estudis Avan\c{c}ats, 08010 Barcelona, Spain}
\newcommand{\QusideAddress}{Quside Technologies S.L., C/Esteve Terradas 1, Of. 217, 08860 Castelldefels (Barcelona), Spain} 

\newcommand{\PRLsection}[1]{\noindent \textit{#1} -- }

\begin{document}

\title{ Bose-Einstein Condensate Comagnetometer}
\author{Pau Gomez}
\affiliation{\ICFOAddress}
\affiliation{\QusideAddress}

\author{Ferran Martin}
\affiliation{\ICFOAddress}
\affiliation{\QusideAddress}

\author{Chiara Mazzinghi}
\affiliation{\ICFOAddress}

\author{Daniel Benedicto Orenes}
\affiliation{\ICFOAddress}

\author{Silvana Palacios}
\affiliation{\ICFOAddress}

\author{Morgan W. Mitchell}
\affiliation{\ICFOAddress}
\affiliation{\ICREAAddress}

\date{\today}

\begin{abstract}
We describe a comagnetometer employing the $f=1$ and $f=2$ ground state hyperfine manifolds of a $^{87}$Rb spinor Bose-Einstein condensate as colocated magnetometers. The hyperfine manifolds feature nearly opposite gyromagnetic ratios and thus the sum of their precession angles is only weakly coupled to external magnetic fields, while being highly sensitive to any effect that rotates both manifolds in the same way. The $f=1$ and $f=2$ transverse magnetizations and azimuth angles are independently measured by nondestructive Faraday rotation probing, and we demonstrate a $\SI{44.0(8)}{\decibel}$ common-mode rejection in good agreement with theory. We show how the magnetometer coherence time can be extended to $\sim\SI{1}{\second}$, by using spin-dependent interactions to inhibit hyperfine relaxing collisions between $f=2$ atoms. The technique could be used in high sensitivity searches for new physics on submillimeter length scales, precision studies of ultracold collision physics, and angle-resolved studies of quantum spin dynamics.
\end{abstract}

\newcommand{\rb}{$^{87}$Rb}

\maketitle
The value of paired magnetic sensors was first demonstrated in the early days of modern magnetism, when C. F. Gauss \cite{Gauss1832, Garland79} used paired compasses to perform the first absolute geomagnetic field measurements. In contemporary physics, paired magnetic sensors enable comagnetometer-based searches for new physics \cite{Weisskopf68, Chupp19}. In a \textit{comagnetometer}, colocated magnetometers respond in the same way to a magnetic field, but have different sensitivities to other, weaker influences. Differential readout then allows high-sensitivity detection of the weak influences with greatly reduced sensitivity to magnetic noise. Comagnetometers have been used to investigate anomalous spin interactions \cite{Vasilakis09, Hunter13, Bulatowicz13, Tullney13, LeePRL2018} and spin-gravity couplings \cite{Venema92, Kimball13, Kimball17} and for stringent tests of Lorentz invariance and CPT violation \cite{Lamoreaux86, Bear00, Cane04, Brown10, Smiciklas11, Allmendinger14}. Further applications are found in inertial navigation and gyroscopes built upon atomic spin comagnetometers \cite{Woodman87,Kornack05, Limes18, Jiang18}. Implementations with miscible mixtures include atomic vapors \cite{Kornack02, Sheng14} and liquid-state NMR with different nuclear spins \cite{Ledbetter12, Wu18}. 

In this Letter, we report a comagnetometer implemented with ultracold atoms, namely a single-domain spinor Bose-Einstein condensate (SBEC). SBECs can have high densities and multisecond magnetic coherence times \cite{Palacios18}, which together imply extreme magnetic sensitivity at the few-\SI{}{\micro\meter} length scale \cite{VengalattorePRL2007}. A single mode SBEC comagnetometer is robust against external magnetic field gradients \cite{VanderbruggenEPL2015} and could find application in detecting short-range spin-dependent forces \cite{Bulatowicz13, Tullney13, LeePRL2018} and studying cold collision physics \cite{Gomez19}. A common limitation in ultracold gas experiments is magnetic field instability, which introduces uncertainty in the Larmor precession. For a typical atomic gyromagnetic ratio of \SI{0.7}{\mega\hertz\per\gauss} and a typical laboratory field fluctuation of $\SI{50}{\micro\gauss}$, the precession angle uncertainty reaches $\SI{\pi/2}{\radian}$ after only a few \SI{}{\milli\second}.   The SBEC comagnetometer overcomes this limitation and resolves coherent phase dynamics at timescales comparable to the lifetime of the ultracold ensemble. 

We employ a $^{87}$Rb SBEC, with the $f=1$ and $f=2$ hyperfine manifolds as colocated magnetic sensors. Because the electron and nuclear spins are anti-aligned (aligned) in the $f=1$ ($f=2$) state, subtraction of the two manifolds' magnetic signals cancels the strong magnetic response -- mostly due to the electron -- while retaining sensitivity to spin-dependent effects that involve the nucleus. The system is well suited to study dipole-dipole \cite{Vasilakis09, Hunter13} and monopole-dipole \cite{Bulatowicz13, Tullney13, LeePRL2018} interactions with ranges down to \SI{\sim10}{\micro\meter}, corresponding to force carriers with masses of up to \SI{\sim 20}{\milli\electronvolt}.  
  A challenge for this strategy is the relatively short lifetime of the $f=2$ manifold produced by exothermic $2\rightarrow 1$ hyperfine-relaxing collisions \cite{Schmaljohann04, Tojo09}.  We strongly suppress these collisions by using the spin-dependent interaction at low magnetic fields to lock the spins in a stretched state.  In this way we achieve $\SI{\sim 1}{\second}$ lifetimes in $f=1,2$ mixtures and a magnetic field noise rejection of $\SI{44.0(8)}{\decibel}$ in the comagnetometer readout.
\\
\\
\PRLsection{Apparatus and state preparation} The comagnetometer is implemented on a superposition of the $f=1,2$ hyperfine manifolds in a single domain SBEC of $^{87}$Rb \cite{Palacios18}. The SBEC is achieved through \SI{4.5}{\second} of all-optical evaporation, reaching a condensate fraction above $90\%$. At the end of evaporation, the potential has a mean trapping frequency $\bar{\omega}=2\pi\times \SI{90(9)}{\hertz}$ \cite{Gomez19} and typically contains $N=N^{(1)}+N^{(2)}\approx \SI{1e5}{atoms}$. 

We work in the single-mode approximation (SMA) \cite{Pu99, Kawaguchi12, Palacios18, Gomez19}, in which the vectorial order parameter $\xi^{(f)}_m$ describes the global spin state. The quantization axis is taken along the magnetic field $\mathbf{B}=B\mathbf{z}$ and the indices label the hyperfine manifolds $f\in\lbrace 1,2\rbrace$ and Zeeman sublevels $m\in\lbrace -f,...,+f\rbrace$. The spin of the system is initialized in the $f=1$ polar state $\xi/\sqrt{N}=(0,1,0)^T\oplus {\bf 0}_5^T$, where the initially empty $f=2$ manifold is denoted by the length-$5$ zero vector ${\bf 0}_5$.

Following the optical evaporation the spin state is prepared in a magnetically sensitive $f=1,2$ superposition. To this purpose, we use microwave (mw) and radio frequency (rf) pulses, coupling the hyperfine manifolds and their Zeeman sublevels, respectively.  First, a rf $\pi/2$ pulse rotates the polar state into $\xi\sqrt{N}=(1/\sqrt{2},0,1/\sqrt{2})^T\oplus {\bf 0}_5^T$. A mw $\pi$ pulse on the $\ket{f=-1,m=-1}\leftrightarrow\ket{f=2,m=-2}$ transition then produces the state $\xi/\sqrt{N}=(1/\sqrt{2},0,0)^T\oplus (0,0,0,0,1/\sqrt{2})^T$, which describes a stretched state oriented along (against) the magnetic field for the $f=1$ ($f=2$) manifold. Finally, both spins are simultaneously rotated into the $\mathbf{x}$-$\mathbf{y}$ plane by means of a second rf $\pi/2$ pulse. 

Note that we use rf fields along the $\mathbf{x}$ or $\mathbf{y}$ directions to simultaneously drive coherent state rotations of the $f=1$ and $f=2$ manifolds. Such fields can be simultaneously resonant due to the nearly opposite gyromagnetic ratios, which we write $\gamma^{(1)}=-\gamma_0-\gamma_s, \; \gamma^{(2)}=+\gamma_0-\gamma_s$, where $\gamma_0/2 \pi \approx \SI{ 700}{\kilo\hertz\gauss^{-1}}$ and $\gamma_s/2\pi\approx \SI{1.39}{\kilo\hertz\gauss^{-1}}$. The rf frequency is tuned to match the Zeeman splitting in $f=1$ and is detuned by $2\gamma_sB<0.12 \Omega$ from the $f=2$ Zeeman splitting, where $\Omega$ is the resonant Rabi frequency.
\\
\\
\PRLsection{Spin evolution and probing} In the transverse plane, the spin manifolds precess around the magnetic field in opposite directions. In the SMA, $f=1$ and $f=2$ experience exactly the same external magnetic field and their angular evolutions read:

\begin{equation}\label{eq:magneticFieldEvolution}
\theta^{(f)}(\mathcal{T})=\int_0^{\mathcal{T}}\gamma^{(f)}B(t)dt\; ,
\end{equation}
where $\theta^{(f)}(\mathcal{T})$ is the azimuthal angle of manifold $f$. The start of the free precession is taken at $t=0$, while its end and start of the readout at $t=\mathcal{T}$.

\newcommand{\supone}{^{(1)}}
\newcommand{\suptwo}{^{(2)}}
\newcommand{\suponetwo}{^{(12)}}
\newcommand{\supf}{^{(f)}}

\begin{figure}[t!]
\centering

\includegraphics[width=1\linewidth]{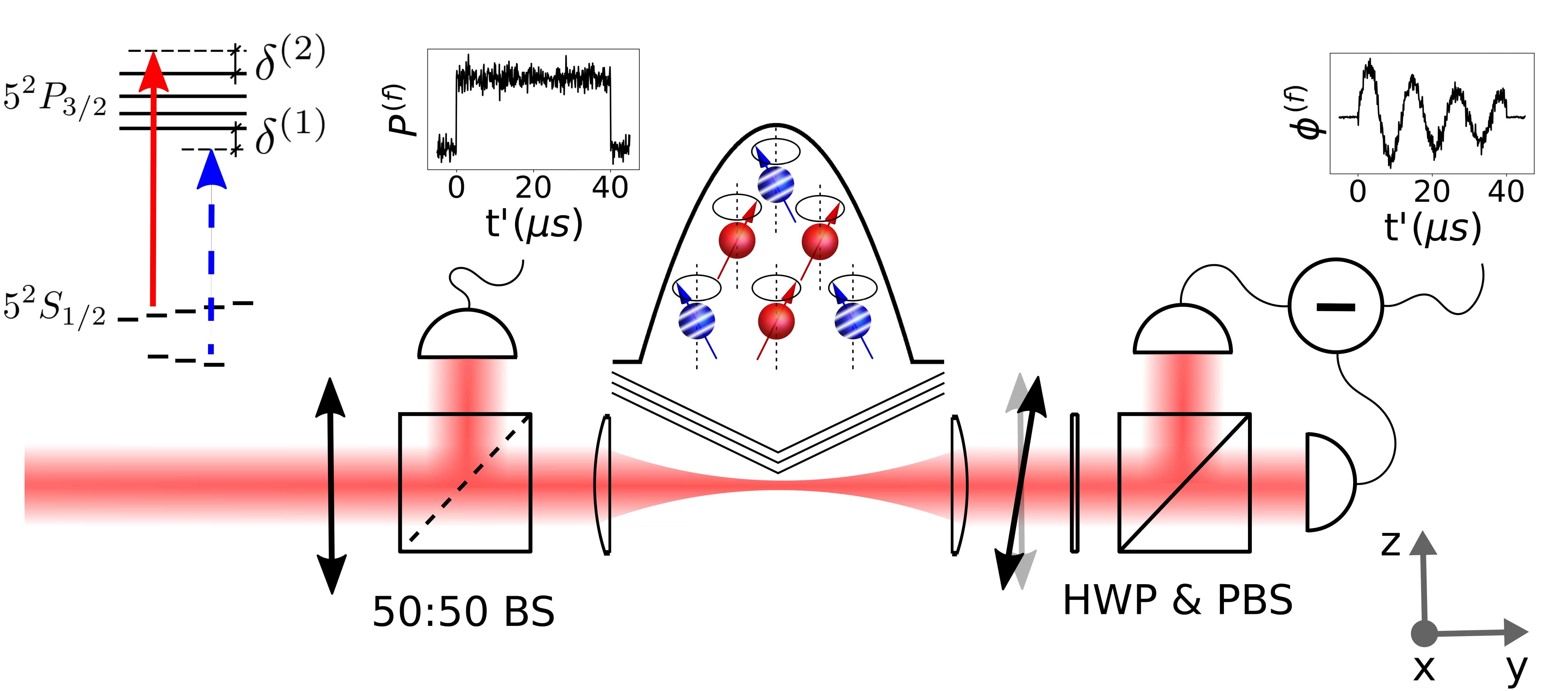} 
\caption{Spin selective Faraday probing of a SBEC in a superposition of $f=1$ (blue, striped) and $f=2$ (red, solid). The frequency of the probe beam alternates between $\delta^{(1)}=\SI{-270}{\mega\hertz}$ red detuned from the $1\leftrightarrow 0'$ transition or $\delta^{(2)}=\SI{360}{\mega\hertz}$ blue detuned from $2\leftrightarrow 3'$ transition ($^{87}$Rb $\mathcal{D}_2$ line), for addressing separately the $f=1$ or $f=2$ manifold. The beam is focused to a few times the Thomas-Fermi radius of the BEC and propagates along the $\mathbf{y}$ direction, while the external magnetic field is applied along $\mathbf{z}$. The linearly polarized probe light experiences a rotation in its polarization proportional to the magnetization along the propagation direction $\phi^{(f)}\propto F_y^{(f)}$. Spin ensembles rotated into the transverse ($\mathbf{x}$-$\mathbf{y}$) plane perform rapid Larmor precessions and the resulting variations in the polarization of the probe light are recorded on a differential photodector. The $\lambda/2$-plate (HWP) and the polarizing beam splitter (PBS) balance and split the orthogonal polarization components before detection. The initial photodetector and 50:50 beam splitter (BS) monitor the power of probe beam $P^{(f)}$. Insets show the acquired signals of $P^{(f)}$ and $\phi^{(f)}$ as a function of probing time $t'$ for a single Faraday readout. Typically, two consecutive \SI{40}{\micro\second} long and \SI{20}{\micro\second} separated Faraday readouts probe $F_\perp^{(f)}$ and $\theta^{(f)}$ in both manifolds. }	
\label{fig:setup}
\end{figure}

The spin state of the ensemble is measured by dispersive Faraday probing \cite{Koschorreck10, Palacios18, Gomez19} as shown in \cref{fig:setup}. We employ linearly polarized probe light closely detuned to the $1\leftrightarrow 0'$ or $2\leftrightarrow 3'$ transitions of the $^{87}$Rb $\mathcal{D}_2$ line, for interrogation of $f=1$ or $f=2$, respectively. The vector atom-light coupling \cite{Geremia06} induces a rotation $\phi^{(f)}$ on the probe polarization, proportional to the atomic spin projection along the propagation direction  ($\mathbf{y}$): $\phi^{(f)}\propto F_y^{(f)}$. The spin projection is written as $F_i^{(f)}\equiv\xi^{(f)\dagger}\hat{F}_i^{(f)}\xi^{(f)}$, where $\hat{F}_i^{(f)}$ are the spin-$f$ matrices along direction $i\in\lbrace x,y,z\rbrace$. The rotation signal is recorded on a balanced differential photodector \cite{Martin16}, from which the evolving spin projection is inferred and is fitted with
\begin{equation}\label{eq:fitFunction}
F_y^{(f)}(t>\mathcal{T})=F_{\perp}^{(f)}(\mathcal{T}) e^{-t'/t_{\rm{dep}}^{(f)}} \sin\left[\gamma^{(f)}\bar{B}t'+\theta^{(f)}(\mathcal{T})\right]
\end{equation}
where the free fit parameters are the transverse spin magnitude $F_\perp^{(f)}(\mathcal{T})$, the azimuth angle $\theta^{(f)}(\mathcal{T})$ and the depolarization rate $1/t_{\rm{dep}}^{(f)}$ due to off-resonant photon scattering. The average magnetic field $\bar{B}$ is calibrated beforehand. In \cref{eq:fitFunction} we distinguish between free evolution time $\mathcal{T}$ and probing time $t'\equiv t-\mathcal{T}$. The first one ranges from tens of $\SI{}{\micro \second}$ to $\SI{1.5}{\second}$, while the second one covers the $\SI{40}{\micro\second}$ of continuous Faraday probing. In the following discussion, we simplify the notation by omitting the explicit $\mathcal{T}$ dependence in the best fit estimates of the transverse spin magnitudes and azimuth angles, writing them as $F_\perp^{(f)}$ and  $\theta^{(f)}$. 

\PRLsection{Comagnetometer}A largely $B$-independent signal is obtained by adding the azimuth estimates to obtain $\theta^{(12)} \equiv \theta^{(1)}+\theta^{(2)}$. We define $\theta^{(12)}$ as our comagnetometer readout. From \cref{eq:magneticFieldEvolution}, its magnetic field contribution is $\theta^{(12)}_B =-2\gamma_s \int_0^{\mathcal{T}} B(t) dt$ and its magnetic field dependency is suppressed by the ratio $|\partial_B\theta^{(f)}/\partial_B\theta^{(12)}| \approx \gamma_0/2\gamma_s  = 251$ (in amplitude) or \SI{48.0}{\decibel} (in power). In contrast, any effect that influences ${\theta}^{(1)}$ and ${\theta}^{(2)}$ in the same direction would doubly influence ${\theta}^{(12)}$. 

\ \\\
\PRLsection{Hyperfine relaxing collisions} The performance of the comagnetometer described above depends strongly on the lifetime imposed by hyperfine relaxing collisions.  In a hyperfine relaxing collision, the liberated energy is transferred to the motional degree of freedom, which expels the colliding atoms from the trap \cite{Tojo09}. This process makes it difficult not only to achieve condensation in $f=2$, but also to observe coherent spinor dynamics in the $f=2$ state and in $f=1,2$ mixtures. 

We divide the hyperfine relaxing collisions into $f=1,2$ collisions ($1\coll 2$) and $f=2,2$  collisions ($2\coll 2$). For the proposed comagnetometer, where $f=1$ and $f=2$ precess in opposite directions, hyperfine relaxing collisions of type $1\coll 2$ are unavoidable and set an upper limit on the lifetime of the ensemble. In contrast, the stronger $2\coll 2$ collisions can be suppressed by preparing $f=2$ in a stretched state, i.e. $|{\bf F^{(2)}}|=2N^{(2)}$.  The stability of stretched spin states is determined by the quadratic Zeeman shift (QZS) and the spin interaction.

The QZS drives coherent orientation-to-alignment oscillations \cite{Palacios18}, e.g. from $ F_\perp\supf = f N\supf$ to $ F_\perp\supf = 0$ and back. In themselves, these oscillations are only a minor inconvenience; they allow full-signal measurements but only at certain times. In combination with the  $2\coll 2$ hyperfine-relaxing collisions, however, the QZS acts to destabilize stretched $f=2$ states and can greatly reduce the $f=2$ lifetime.

The ferromagnetic (antiferromagnetic) spin interaction in $f=1$ ($f=2$) \cite{Kawaguchi12, Saito18}, which lowers (raises) the energy of stretched states relative to other states, opposes the orientation-to-alignment conversion and can reestablish long $f=2$ lifetimes. 

The competition of QZS and spin interaction effects is parametrized by the ratio $\eta\supf \equiv |{E}_q^{(f)}/E\subint^{(f)}|$,  where the QZS and spin interaction energies of a transverse stretched state in hyperfine manifold $f$ are 
\begin{subequations} \label{eq:energies}
\begin{align}
\label{eq:energiesQZS}
E_q\supf&=(-1)^{f-1}\frac{\left(\hbar\gamma\supf B\right)^2}{\hbar\omega\subhfs}\frac{fN\supf}{2} \; ,\\
\label{eq:energiesINT}
E\subint\supf&=\frac{g_1\supf}{2V\subeff}\left(fN\supf\right)^2.
\end{align}
\end{subequations}
\\
Here $\hbar$ is the Planck constant, $\omega\subhfs=2\pi\times\SI{6.8}{\giga\hertz}$ is the $f=1,2$ hyperfine splitting frequency and the spin interaction coefficients $g_1^{(f)}$ and effective volume $V\subeff$ are defined in \cite{Gomez19}. When $\eta\supf \ll 1$ the orientation-to-alignment oscillations are suppressed, which prevents $2\coll 2$ hyperfine-relaxing collisions in initially stretched  $f=2$ states.

\begin{figure}[t!]
\centering
\includegraphics[width=1\linewidth]{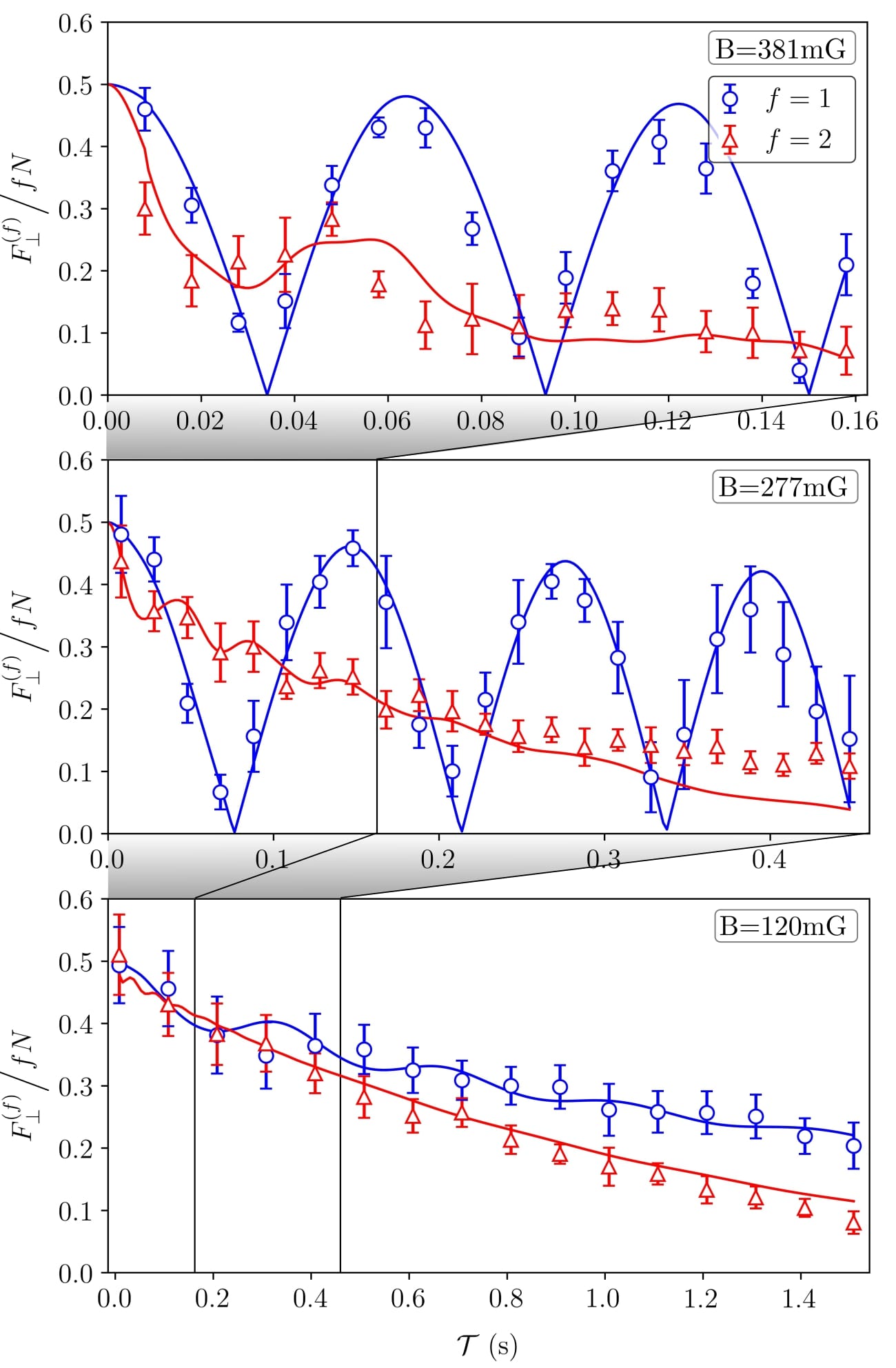} 
\caption{Evolution of the comagnetometer transverse spin magnitude for $f=1$ (blue circles) and $f=2$ (red triangles) for decreasing magnetic field strengths $B$. Graphs show the transverse degree of polarization $F_{\perp}^{(f)}/(f N)$ versus evolution time $\mathcal{T}$, where the atom number $N=2F^{(1)}_\perp(\mathcal{T}=0)$ is estimated from the first $f=1$ Faraday rotation signal. At field strengths $\textrm{B}= \lbrace\SI{381}{\milli \gauss},\SI{277}{\milli \gauss},\SI{120}{\milli \gauss} \rbrace$ we have $N=\lbrace1.47(11),1.05(13),1.15(14) \rbrace \times 10^5$ atoms and ratios between QZS and spin interaction energies of $\eta\supone = \lbrace 5.62,3.40,0.62 \rbrace$ and $\eta\suptwo =  \lbrace 1.01, 0.61,0.11\rbrace$, respectively. The reduction in $F_\perp^{(2)}$ results from hyperfine relaxing collisions throughout the evolution time, which is constrained to $\mathcal{T}\geq\SI{4}{\milli \second}+\SI{4}{\milli \second}$ by the magnetic ramps at the beginning and end of the experimental sequence. Solid lines are SMA mean-field simulations as described in the text.  Error bars show the measured standard deviation in the transverse spin magnitude over 12 experimental repetitions and black vertical lines indicate the temporal extent of the graphs above.}	
\label{fig:lifetimeCoMag}
\end{figure}

In \cref{fig:lifetimeCoMag} we show results on orientation-to-alignment oscillations and hyperfine-relaxing relaxation for different applied magnetic fields.  The state preparation is performed at $B = \SI{282}{\milli \gauss}$ and, as described above, results in a superposition of transversely stretched states $\xi/\sqrt{N}=\hat{R}_{\frac{\pi}{2}}^{(1)}(1/\sqrt{2},0,0)^T\oplus\hat{R}_{\frac{\pi}{2}}^{(2)} (0,0,0,0,1/\sqrt{2})^T$, where $\hat{R}_{\frac{\pi}{2}}^{(f)}$ denotes the rf $\pi/2$ rotation into the transverse plane. Thereafter, the magnetic field is ramped in \SI{4}{\milli \second} to a value of \SI{381}{\milli \gauss}, \SI{277}{\milli \gauss} or \SI{120}{\milli \gauss} for free evolution. In the \SI{4}{\milli \second} prior to Faraday readout, the field is ramped back to \SI{282}{\milli \gauss} to have a consistent readout process. 

We observe clear orientation to alignment conversion cycles in $f=1$ at $\SI{381}{\milli \gauss}$ and $\SI{277}{\milli \gauss}$. The oscillatory process is less visible in $f=2$ due to its stronger spin interaction and rapid atom losses via $2\coll 2$ hyperfine relaxing collisions. At \SI{120}{\milli \gauss}, $(\eta\supone,\eta\suptwo) = (0.62, 0.11 ) \ll 1$ and the spin interaction dominates in both hyperfine manifolds. As a result, $2\coll 2$ losses are suppressed and the $\sim\SI{1}{\second}$ lifetime is limited by $1\coll 2$ hyperfine relaxing collisions. 
 
\PRLsection{Modeling}
We use SMA mean field simulations including intra- and interhyperfine interactions \cite{Gomez19}, with two-body loss channels included as $a_\mathcal{C}^{(2)}\rightarrow a_\mathcal{C}^{(2)} - i \tilde{a}^{(2)}_\mathcal{C}$ and $a_\mathcal{C}^{(12)}\rightarrow a_\mathcal{C}^{(12)} - i \tilde{a}^{(12)}_\mathcal{C}$, where $\mathcal{C}$ is the total spin of a given collision channel \cite{Tojo09}. A full set of scattering rates is not known, so for simplicity we take  $\tilde{a}^{(2)}_\mathcal{C} = \tilde{a}^{(2)}=\SI{0.692(34)}{\bohr}$ and $\tilde{a}^{(12)}_\mathcal{C} = \tilde{a}^{(12)}=\SI{0.0110(11)}{\bohr}$, values found from fitting  the Faraday rotation signals of, respectively, $f=2$ at \SI{381}{\milli\gauss} (upper panel of Fig.~\ref{fig:lifetimeCoMag}) and $f=1$ at \SI{120}{\milli\gauss} (lower panel of Fig.~\ref{fig:lifetimeCoMag}).  

\PRLsection{Magnetic background suppression} We proceed by evaluating the comagnetometer common-mode suppression at low magnetic fields, where both hyperfine manifolds are long lived. To this end, a constant bias magnetic field of $\SI{120}{\milli\gauss}$ is applied for state preparation, hold time and Faraday readout. This removes the temporal overhead of the previously required magnetic ramps such that hold times down to $\SI{20}{\micro\second}$ are accessible, limited only by the hardware timing of the experiment. 

\begin{figure}[t]
\centering\includegraphics[width=1\linewidth]{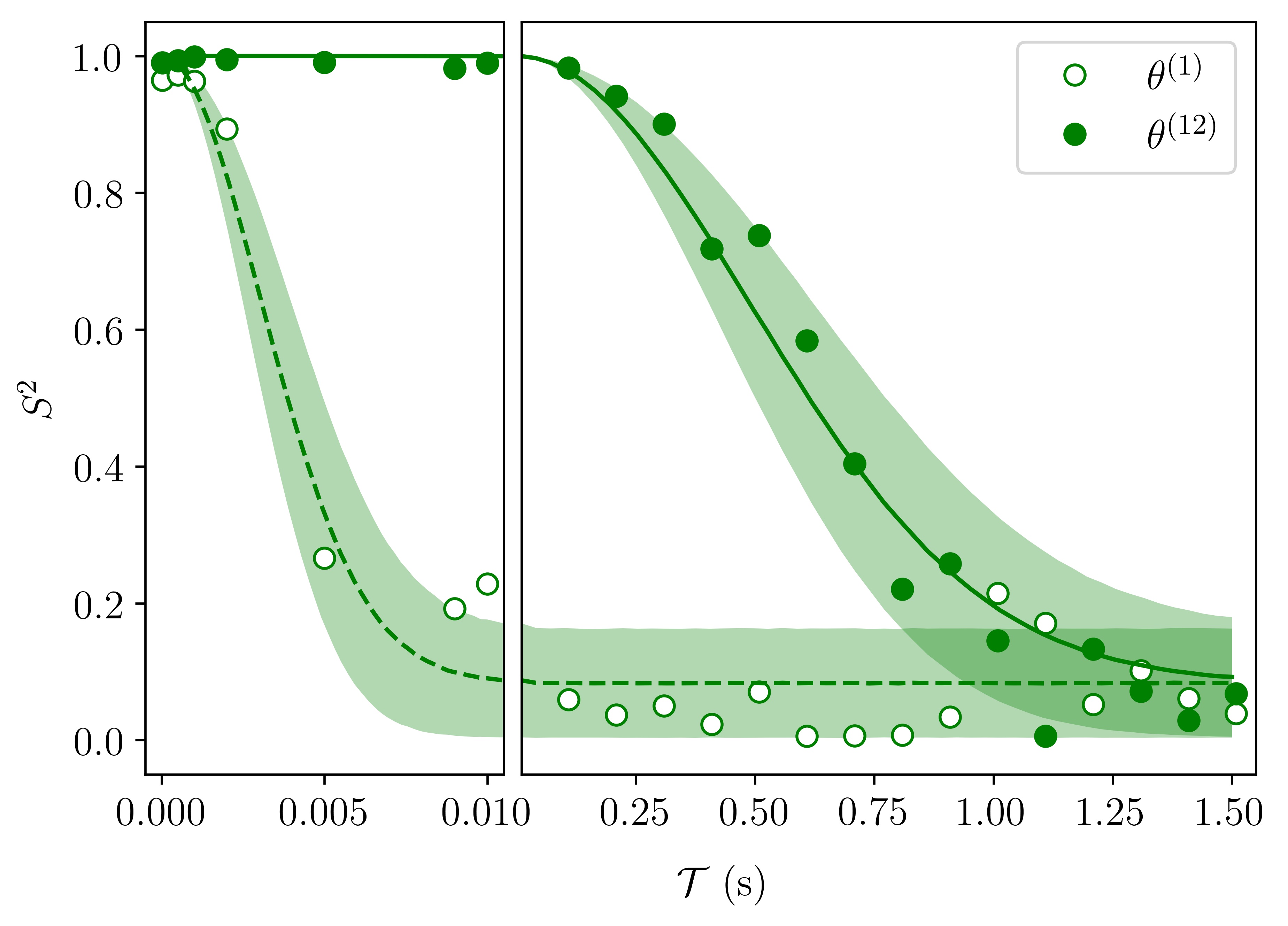} 
\caption{Magnetic noise rejection of the SBEC comagnetometer. Graph shows evolution of $S^2$, where $S \equiv |\langle \exp[i \theta] \rangle |$ is the sharpness, as a function of hold time ${\cal T}$, for $\theta = \theta\supone$ and $\theta = \theta\suponetwo$.  Averages are taken over $K=12$ samples. Sharpness of $\theta\suptwo$ closely tracks that of  $\theta\supone$ and is not shown. Dashed and solid lines show fits assuming a quasistatic field noise model [see  \cref{eq:R} and surrounding paragraph], yielding rms deviations in the azimuth angles and comagnetometer readout of $\sigma_{\theta^{(1)}}= \SI{230(20)}{\radian\per\second}\mathcal{T}$ and $\sigma_{\theta^{(12)}}= \SI{1.45(5)}{\radian\per\second}\mathcal{T}$, respectively. Shaded areas represent the numerically estimated plus and minus one standard deviation in $S^2$ for the above obtained fit results.	}
\label{fig:magNoise}
\end{figure}

We measure the spread in estimated azimuth angles $\theta^{(f)}$ and comagnetometer signal $\theta^{(12)}$ as a function of hold time $\mathcal{T}$, with results shown in \cref{fig:magNoise}. We employ as a cyclic statistic the \textit{sharpness} $S \equiv |\langle \exp[i {\theta}] \rangle|$ \cite{Berry01}, where $\langle \cdot \rangle$ here indicates the sample mean and $\theta$ is an angle variable, e.g. $\theta^{(f)}$ or $\theta^{(12)}$. 
$S^2=1$ indicates no spread of $\theta$ while $S^2$ near zero indicates a large spread. 
We can relate the loss of sharpness with increasing $\mathcal{T}$ seen in \cref{fig:magNoise} to the magnetic noise as follows.  First we note that the hold time $\mathcal{T}$ is always small relative to the time between measurements and that by \cref{eq:magneticFieldEvolution}, $\theta^{(f)}$ is most sensitive to the dc component of $B(t)$.  This motivates a quasistatic model, where the field $B$ is constant during free evolution and normally distributed from shot to shot, with variance $\sigma_B^2$.  Consequently $\theta^{(f)}$ and $\theta^{(12)}$ are normally distributed, with rms deviations $\sigma_{\theta^{(f)}}=|\gamma^{(f)}|\sigma_B \mathcal{T} \approx\gamma_0\sigma_B \mathcal{T}$  and $\sigma_{\theta^{(12)}}=2\gamma_s\sigma_B \mathcal{T}$.  
For normally distributed $\theta$ and sample size $K$, the expectation of $S^2$ is
\begin{equation}\label{eq:R}
\langle S^2 \rangle=\frac{1}{K}+\frac{K-1}{K}e^{-\sigma_\theta^2}. 
\end{equation}
This form is fitted to the data of \cref{fig:magNoise} to find $\sigma_{\theta^{(1)}}=  \SI{230(20)}{\radian\per\second}\mathcal{T}$ and $\sigma_{\theta^{(12)}}= \SI{1.45(5)}{\radian\per\second}\mathcal{T}$. 

The ratio between these indicates a common-mode rejection of $B$ fluctuations $|\partial_B \theta^{(1)}/\partial_B \theta^{(12)}| = 159(15)$ in amplitude or $\SI{44.0(8)}{\decibel}$ in power, in reasonable agreement with the predicted \SI{48}{\decibel} rejection.  The discrepancy is plausibly due to field drifts during the free evolution, which principally affect larger $\mathcal{T}$ and thus $\sigma_{\theta^{(12)}}$. 

\PRLsection{Conclusions and outlook} We have presented a SBEC comagnetometer implemented on a superposition of stretched states in the $f=1$ and $f=2$  ground state hyperfine manifolds of $^{87}$Rb. Hyperfine relaxing collisions among $f=2$ atoms are suppressed by operating the system at low magnetic fields, where the spin interaction energy dominates over the QZS. The observed coherent spin dynamics and atom losses are in good agreement with SMA mean field simulations. We demonstrate a \SI{44.0(8)}{\decibel} reduction in sensitivity to magnetic fields, while retaining sensitivity to effects that rotate both hyperfine ground states in the same way. 

This comagnetometer has already been used for precision measurement of interhyperfine interactions in ultracold gases \cite{Gomez19} and could be used to detect exotic spin couplings.  The signal $\theta_1 + \theta_2$ is largely insensitive to $B$, which couples principally to the electron spin, but is sensitive to any effect that couples to principally to the nuclear spin, or indeed to the electron and nuclear spins with a ratio different than that of the magnetic coupling.

The  equivalent magnetic sensitivity is $\delta B^{(12)} \approx t_{\rm cyc}^{1/2}(\gamma_0 t_{\rm coh})^{-1} \sqrt{ (\delta \theta\supone)^2 + (\delta \theta\suptwo)^2}$, where $t_{\rm coh}$ and $t_{\rm cyc}$ are the coherence and cycle times, respectively, and $\delta \theta\supf$ is the readout uncertainty of $\theta\supf$.   In the present implementation  $t_{\rm coh} \approx \SI{1}{\second}$, such that for evolution time $\mathcal{T}\approx t_{\rm coh}$ we typically have $N\supf\approx 2\times 10^4$.  For a cycling time of $t_{\rm cyc} = t_{\rm coh} + \SI{19}{\second}$ and a readout noise of 1000 spins ($\delta \theta\supf \approx 1000/N\supf$) this gives $\delta B^{(12)} \approx \SI{8}{\pico\tesla\per\sqrt\hertz}$.

We note a few natural extensions of the technique. First, the remaining QZS can be cancelled using microwave dressing, to allow free choice of Larmor frequency and zero hyperfine relaxing collisions between $f=2$ atoms. Second, a state-specific optical Zeeman shift can be applied to null $\gamma_s$ and thus fully cancel background field noise.  Third, a softer confining potential could reduce the rate of $1\coll 2$ collisions, to give $\tau^{(12)} \approx \SI{8}{\second}$ if $\bar{\omega}\approx 2 \pi \times \SI{30}{\hertz}$. Cavity-assisted readout  \cite{LodewyckPRA2009, Schleier-SmithPRL2010, LeeOL2014} could be used to reach the projection-noise level $\delta \theta\supf = 1/\sqrt{2 f N\supf}$ while faster loading could give $t_{\rm cyc} \approx t_{\rm coh} \approx \tau^{(12)}$.  Combining these would give a sensitivity  $\delta B^{(12)}\approx \SI{50}{\femto\tesla\per\sqrt\hertz}$ or $\delta E/h \approx \SI{360}{\micro\hertz\per\sqrt\hertz}$, where $\delta E$ is the sensitivity on a hyperfine dependent energy splitting.

In one week of running time, the statistical uncertainty of such a system would reach $\approx \SI{65}{\atto\tesla}$, comparable to state-of-the-art vapor- and gas-phase comagnetometers used in searches for physics beyond the standard model.  For example, Lee\textit{ et al.} report \SI{70}{\atto\tesla} residual uncertainty after \SI{1.5}{week} of acquisition in a recent search for axion-like particles with a $^3$He-K comagnetometer \cite{LeePRL2018}. A SBEC comagnetometer would moreover be able to probe length scales down to $\SI{\sim 10}{\micro\meter}$, about four orders of magnitude shorter than other comagnetometers. In searches for axion-like particles, these length scales are only weakly constrained by astrophysical arguments \cite{RaffeltPRD2012} and prior laboratory tests \cite{PetukhovPRL2010, SerebrovJETPL2010}.

Another potential application is angle-resolved spin amplification.  Spin amplifiers use coherent collision processes in a BEC to achieve high-gain, quantum-noise limited amplification of small spin perturbations \cite{Leslie09}. They are of particular interest in studies of quantum dynamics and nonclassical state generation \cite{Klempt10}, but to date have not been able to resolve the magnetically sensitive azimuthal spin degree of freedom. This issue can be circumvented in a SBEC comagnetometer in which one hyperfine manifold tracks the magnetic field evolution while the other experiences parametric spin amplification. 

\PRLsection{Acknowledgements}
We thank D. Budker and M. Romalis for helpful discussions. This project was supported by Spanish MINECO projects MAQRO (Grant No. FIS2015-68039-P), OCARINA (Grant No. PGC2018-097056-B-I00) and Q-CLOCKS (Grant No. PCI2018-092973), the Severo Ochoa program (Grant No. SEV-2015-0522); Ag\`{e}ncia de Gesti\'{o} d'Ajuts Universitaris i de Recerca (AGAUR) project (Grant No. 2017-SGR-1354); Fundaci\'{o} Privada Cellex and Generalitat de Catalunya (CERCA program); Quantum Technology Flagship project MACQSIMAL (Grant No. 820393); Marie Sk{\l{odowska-Curie ITN ZULF-NMR (Grant No. 766402); 17FUN03-USOQS, which has received funding from the EMPIR programme cofinanced by the Participating States and from the European Union's Horizon 2020 research and innovation program.


\bibliographystyle{apsrev4-1no-url}
\bibliography{../Biblio/biblio}{}

\begin{thebibliography}{47}%
\makeatletter
\providecommand \@ifxundefined [1]{%
 \@ifx{#1\undefined}
}%
\providecommand \@ifnum [1]{%
 \ifnum #1\expandafter \@firstoftwo
 \else \expandafter \@secondoftwo
 \fi
}%
\providecommand \@ifx [1]{%
 \ifx #1\expandafter \@firstoftwo
 \else \expandafter \@secondoftwo
 \fi
}%
\providecommand \natexlab [1]{#1}%
\providecommand \enquote  [1]{``#1''}%
\providecommand \bibnamefont  [1]{#1}%
\providecommand \bibfnamefont [1]{#1}%
\providecommand \citenamefont [1]{#1}%
\providecommand \href@noop [0]{\@secondoftwo}%
\providecommand \href [0]{\begingroup \@sanitize@url \@href}%
\providecommand \@href[1]{\@@startlink{#1}\@@href}%
\providecommand \@@href[1]{\endgroup#1\@@endlink}%
\providecommand \@sanitize@url [0]{\catcode `\\12\catcode `\$12\catcode
  `\&12\catcode `\#12\catcode `\^12\catcode `\_12\catcode `\%12\relax}%
\providecommand \@@startlink[1]{}%
\providecommand \@@endlink[0]{}%
\providecommand \url  [0]{\begingroup\@sanitize@url \@url }%
\providecommand \@url [1]{\endgroup\@href {#1}{\urlprefix }}%
\providecommand \urlprefix  [0]{URL }%
\providecommand \Eprint [0]{\href }%
\providecommand \doibase [0]{http://dx.doi.org/}%
\providecommand \selectlanguage [0]{\@gobble}%
\providecommand \bibinfo  [0]{\@secondoftwo}%
\providecommand \bibfield  [0]{\@secondoftwo}%
\providecommand \translation [1]{[#1]}%
\providecommand \BibitemOpen [0]{}%
\providecommand \bibitemStop [0]{}%
\providecommand \bibitemNoStop [0]{.\EOS\space}%
\providecommand \EOS [0]{\spacefactor3000\relax}%
\providecommand \BibitemShut  [1]{\csname bibitem#1\endcsname}%
\let\auto@bib@innerbib\@empty
\bibitem [{\citenamefont {Gau{\ss}}(1832)}]{Gauss1832}%
  \BibitemOpen
  \bibfield  {author} {\bibinfo {author} {\bibfnamefont {C.~F.}\ \bibnamefont
  {Gau{\ss}}},\ }\href@noop {} {\bibfield  {journal} {\bibinfo  {journal}
  {Intensitas vis magneticae terrestris ad mensuram absolutam revocata}\ }
  (\bibinfo {year} {1832})}\BibitemShut {NoStop}%
\bibitem [{\citenamefont {Garland}(1979)}]{Garland79}%
  \BibitemOpen
  \bibfield  {author} {\bibinfo {author} {\bibfnamefont {G.~D.}\ \bibnamefont
  {Garland}},\ }\href@noop {} {\bibfield  {journal} {\bibinfo  {journal}
  {Historia Mathematica}\ }\textbf {\bibinfo {volume} {6}},\ \bibinfo {pages}
  {5} (\bibinfo {year} {1979})}\BibitemShut {NoStop}%
\bibitem [{\citenamefont {Weisskopf}\ \textit {et~al.}(1968)\citenamefont
  {Weisskopf}, \citenamefont {Carrico}, \citenamefont {Gould}, \citenamefont
  {Lipworth},\ and\ \citenamefont {Stein}}]{Weisskopf68}%
  \BibitemOpen
  \bibfield  {author} {\bibinfo {author} {\bibfnamefont {M.~C.}\ \bibnamefont
  {Weisskopf}}, \bibinfo {author} {\bibfnamefont {J.~P.}\ \bibnamefont
  {Carrico}}, \bibinfo {author} {\bibfnamefont {H.}~\bibnamefont {Gould}},
  \bibinfo {author} {\bibfnamefont {E.}~\bibnamefont {Lipworth}}, \ and\
  \bibinfo {author} {\bibfnamefont {T.~S.}\ \bibnamefont {Stein}},\ }\href
  {\doibase10.1103/PhysRevLett.21.1645} {\bibfield  {journal} {\bibinfo
  {journal} {Phys. Rev. Lett.}\ }\textbf {\bibinfo {volume} {21}},\ \bibinfo
  {pages} {1645} (\bibinfo {year} {1968})}\BibitemShut {NoStop}%
\bibitem [{\citenamefont {Chupp}\ \textit {et~al.}(2019)\citenamefont {Chupp},
  \citenamefont {Fierlinger}, \citenamefont {Ramsey-Musolf},\ and\
  \citenamefont {Singh}}]{Chupp19}%
  \BibitemOpen
  \bibfield  {author} {\bibinfo {author} {\bibfnamefont {T.~E.}\ \bibnamefont
  {Chupp}}, \bibinfo {author} {\bibfnamefont {P.}~\bibnamefont {Fierlinger}},
  \bibinfo {author} {\bibfnamefont {M.~J.}\ \bibnamefont {Ramsey-Musolf}}, \
  and\ \bibinfo {author} {\bibfnamefont {J.~T.}\ \bibnamefont {Singh}},\ }\href
  {\doibase10.1103/RevModPhys.91.015001} {\bibfield  {journal} {\bibinfo
  {journal} {Rev. Mod. Phys.}\ }\textbf {\bibinfo {volume} {91}},\ \bibinfo
  {pages} {015001} (\bibinfo {year} {2019})}\BibitemShut {NoStop}%
\bibitem [{\citenamefont {Vasilakis}\ \textit {et~al.}(2009)\citenamefont
  {Vasilakis}, \citenamefont {Brown}, \citenamefont {Kornack},\ and\
  \citenamefont {Romalis}}]{Vasilakis09}%
  \BibitemOpen
  \bibfield  {author} {\bibinfo {author} {\bibfnamefont {G.}~\bibnamefont
  {Vasilakis}}, \bibinfo {author} {\bibfnamefont {J.~M.}\ \bibnamefont
  {Brown}}, \bibinfo {author} {\bibfnamefont {T.~W.}\ \bibnamefont {Kornack}},
  \ and\ \bibinfo {author} {\bibfnamefont {M.~V.}\ \bibnamefont {Romalis}},\
  }\href {\doibase10.1103/PhysRevLett.103.261801} {\bibfield  {journal}
  {\bibinfo  {journal} {Phys. Rev. Lett.}\ }\textbf {\bibinfo {volume} {103}},\
  \bibinfo {pages} {261801} (\bibinfo {year} {2009})}\BibitemShut {NoStop}%
\bibitem [{\citenamefont {Hunter}\ \textit {et~al.}(2013)\citenamefont
  {Hunter}, \citenamefont {Gordon}, \citenamefont {Peck}, \citenamefont {Ang},\
  and\ \citenamefont {Lin}}]{Hunter13}%
  \BibitemOpen
  \bibfield  {author} {\bibinfo {author} {\bibfnamefont {L.}~\bibnamefont
  {Hunter}}, \bibinfo {author} {\bibfnamefont {J.}~\bibnamefont {Gordon}},
  \bibinfo {author} {\bibfnamefont {S.}~\bibnamefont {Peck}}, \bibinfo {author}
  {\bibfnamefont {D.}~\bibnamefont {Ang}}, \ and\ \bibinfo {author}
  {\bibfnamefont {J.-F.}\ \bibnamefont {Lin}},\ }\href
  {\doibase10.1126/science.1227460} {\bibfield  {journal} {\bibinfo  {journal}
  {Science}\ }\textbf {\bibinfo {volume} {339}},\ \bibinfo {pages} {928}
  (\bibinfo {year} {2013})}\BibitemShut {NoStop}%
\bibitem [{\citenamefont {Bulatowicz}\ \textit {et~al.}(2013)\citenamefont
  {Bulatowicz}, \citenamefont {Griffith}, \citenamefont {Larsen}, \citenamefont
  {Mirijanian}, \citenamefont {Fu}, \citenamefont {Smith}, \citenamefont
  {Snow}, \citenamefont {Yan},\ and\ \citenamefont {Walker}}]{Bulatowicz13}%
  \BibitemOpen
  \bibfield  {author} {\bibinfo {author} {\bibfnamefont {M.}~\bibnamefont
  {Bulatowicz}}, \bibinfo {author} {\bibfnamefont {R.}~\bibnamefont
  {Griffith}}, \bibinfo {author} {\bibfnamefont {M.}~\bibnamefont {Larsen}},
  \bibinfo {author} {\bibfnamefont {J.}~\bibnamefont {Mirijanian}}, \bibinfo
  {author} {\bibfnamefont {C.~B.}\ \bibnamefont {Fu}}, \bibinfo {author}
  {\bibfnamefont {E.}~\bibnamefont {Smith}}, \bibinfo {author} {\bibfnamefont
  {W.~M.}\ \bibnamefont {Snow}}, \bibinfo {author} {\bibfnamefont
  {H.}~\bibnamefont {Yan}}, \ and\ \bibinfo {author} {\bibfnamefont {T.~G.}\
  \bibnamefont {Walker}},\ }\href {\doibase10.1103/PhysRevLett.111.102001}
  {\bibfield  {journal} {\bibinfo  {journal} {Phys. Rev. Lett.}\ }\textbf
  {\bibinfo {volume} {111}},\ \bibinfo {pages} {102001} (\bibinfo {year}
  {2013})}\BibitemShut {NoStop}%
\bibitem [{\citenamefont {Tullney}\ \textit {et~al.}(2013)\citenamefont
  {Tullney}, \citenamefont {Allmendinger}, \citenamefont {Burghoff},
  \citenamefont {Heil}, \citenamefont {Karpuk}, \citenamefont {Kilian},
  \citenamefont {Knappe-Gr\"uneberg}, \citenamefont {M\"uller}, \citenamefont
  {Schmidt}, \citenamefont {Schnabel}, \citenamefont {Seifert}, \citenamefont
  {Sobolev},\ and\ \citenamefont {Trahms}}]{Tullney13}%
  \BibitemOpen
  \bibfield  {author} {\bibinfo {author} {\bibfnamefont {K.}~\bibnamefont
  {Tullney}}, \bibinfo {author} {\bibfnamefont {F.}~\bibnamefont
  {Allmendinger}}, \bibinfo {author} {\bibfnamefont {M.}~\bibnamefont
  {Burghoff}}, \bibinfo {author} {\bibfnamefont {W.}~\bibnamefont {Heil}},
  \bibinfo {author} {\bibfnamefont {S.}~\bibnamefont {Karpuk}}, \bibinfo
  {author} {\bibfnamefont {W.}~\bibnamefont {Kilian}}, \bibinfo {author}
  {\bibfnamefont {S.}~\bibnamefont {Knappe-Gr\"uneberg}}, \bibinfo {author}
  {\bibfnamefont {W.}~\bibnamefont {M\"uller}}, \bibinfo {author}
  {\bibfnamefont {U.}~\bibnamefont {Schmidt}}, \bibinfo {author} {\bibfnamefont
  {A.}~\bibnamefont {Schnabel}}, \bibinfo {author} {\bibfnamefont
  {F.}~\bibnamefont {Seifert}}, \bibinfo {author} {\bibfnamefont
  {Y.}~\bibnamefont {Sobolev}}, \ and\ \bibinfo {author} {\bibfnamefont
  {L.}~\bibnamefont {Trahms}},\ }\href {\doibase10.1103/PhysRevLett.111.100801}
  {\bibfield  {journal} {\bibinfo  {journal} {Phys. Rev. Lett.}\ }\textbf
  {\bibinfo {volume} {111}},\ \bibinfo {pages} {100801} (\bibinfo {year}
  {2013})}\BibitemShut {NoStop}%
\bibitem [{\citenamefont {Lee}\ \textit {et~al.}(2018)\citenamefont {Lee},
  \citenamefont {Almasi},\ and\ \citenamefont {Romalis}}]{LeePRL2018}%
  \BibitemOpen
  \bibfield  {author} {\bibinfo {author} {\bibfnamefont {J.}~\bibnamefont
  {Lee}}, \bibinfo {author} {\bibfnamefont {A.}~\bibnamefont {Almasi}}, \ and\
  \bibinfo {author} {\bibfnamefont {M.}~\bibnamefont {Romalis}},\ }\href
  {\doibase10.1103/PhysRevLett.120.161801} {\bibfield  {journal} {\bibinfo
  {journal} {Phys. Rev. Lett.}\ }\textbf {\bibinfo {volume} {120}},\ \bibinfo
  {pages} {161801} (\bibinfo {year} {2018})}\BibitemShut {NoStop}%
\bibitem [{\citenamefont {Venema}\ \textit {et~al.}(1992)\citenamefont
  {Venema}, \citenamefont {Majumder}, \citenamefont {Lamoreaux}, \citenamefont
  {Heckel},\ and\ \citenamefont {Fortson}}]{Venema92}%
  \BibitemOpen
  \bibfield  {author} {\bibinfo {author} {\bibfnamefont {B.~J.}\ \bibnamefont
  {Venema}}, \bibinfo {author} {\bibfnamefont {P.~K.}\ \bibnamefont
  {Majumder}}, \bibinfo {author} {\bibfnamefont {S.~K.}\ \bibnamefont
  {Lamoreaux}}, \bibinfo {author} {\bibfnamefont {B.~R.}\ \bibnamefont
  {Heckel}}, \ and\ \bibinfo {author} {\bibfnamefont {E.~N.}\ \bibnamefont
  {Fortson}},\ }\href {\doibase10.1103/PhysRevLett.68.135} {\bibfield
  {journal} {\bibinfo  {journal} {Phys. Rev. Lett.}\ }\textbf {\bibinfo
  {volume} {68}},\ \bibinfo {pages} {135} (\bibinfo {year} {1992})}\BibitemShut
  {NoStop}%
\bibitem [{\citenamefont {Kimball}\ \textit {et~al.}(2013)\citenamefont
  {Kimball}, \citenamefont {Lacey}, \citenamefont {Valdez}, \citenamefont
  {Swiatlowski}, \citenamefont {Rios}, \citenamefont {Peregrina-Ramirez},
  \citenamefont {Montcrieffe}, \citenamefont {Kremer}, \citenamefont {Dudley},\
  and\ \citenamefont {Sanchez}}]{Kimball13}%
  \BibitemOpen
  \bibfield  {author} {\bibinfo {author} {\bibfnamefont {D.~F.~J.}\
  \bibnamefont {Kimball}}, \bibinfo {author} {\bibfnamefont {I.}~\bibnamefont
  {Lacey}}, \bibinfo {author} {\bibfnamefont {J.}~\bibnamefont {Valdez}},
  \bibinfo {author} {\bibfnamefont {J.}~\bibnamefont {Swiatlowski}}, \bibinfo
  {author} {\bibfnamefont {C.}~\bibnamefont {Rios}}, \bibinfo {author}
  {\bibfnamefont {R.}~\bibnamefont {Peregrina-Ramirez}}, \bibinfo {author}
  {\bibfnamefont {C.}~\bibnamefont {Montcrieffe}}, \bibinfo {author}
  {\bibfnamefont {J.}~\bibnamefont {Kremer}}, \bibinfo {author} {\bibfnamefont
  {J.}~\bibnamefont {Dudley}}, \ and\ \bibinfo {author} {\bibfnamefont
  {C.}~\bibnamefont {Sanchez}},\ }\href {\doibase10.1002/andp.201300036}
  {\bibfield  {journal} {\bibinfo  {journal} {Annalen der Physik}\ }\textbf
  {\bibinfo {volume} {525}},\ \bibinfo {pages} {514} (\bibinfo {year}
  {2013})}\BibitemShut {NoStop}%
\bibitem [{\citenamefont {Jackson~Kimball}\ \textit {et~al.}(2017)\citenamefont
  {Jackson~Kimball}, \citenamefont {Dudley}, \citenamefont {Li}, \citenamefont
  {Patel},\ and\ \citenamefont {Valdez}}]{Kimball17}%
  \BibitemOpen
  \bibfield  {author} {\bibinfo {author} {\bibfnamefont {D.~F.}\ \bibnamefont
  {Jackson~Kimball}}, \bibinfo {author} {\bibfnamefont {J.}~\bibnamefont
  {Dudley}}, \bibinfo {author} {\bibfnamefont {Y.}~\bibnamefont {Li}}, \bibinfo
  {author} {\bibfnamefont {D.}~\bibnamefont {Patel}}, \ and\ \bibinfo {author}
  {\bibfnamefont {J.}~\bibnamefont {Valdez}},\ }\href
  {\doibase10.1103/PhysRevD.96.075004} {\bibfield  {journal} {\bibinfo
  {journal} {Phys. Rev. D}\ }\textbf {\bibinfo {volume} {96}},\ \bibinfo
  {pages} {075004} (\bibinfo {year} {2017})}\BibitemShut {NoStop}%
\bibitem [{\citenamefont {Lamoreaux}\ \textit {et~al.}(1986)\citenamefont
  {Lamoreaux}, \citenamefont {Jacobs}, \citenamefont {Heckel}, \citenamefont
  {Raab},\ and\ \citenamefont {Fortson}}]{Lamoreaux86}%
  \BibitemOpen
  \bibfield  {author} {\bibinfo {author} {\bibfnamefont {S.~K.}\ \bibnamefont
  {Lamoreaux}}, \bibinfo {author} {\bibfnamefont {J.~P.}\ \bibnamefont
  {Jacobs}}, \bibinfo {author} {\bibfnamefont {B.~R.}\ \bibnamefont {Heckel}},
  \bibinfo {author} {\bibfnamefont {F.~J.}\ \bibnamefont {Raab}}, \ and\
  \bibinfo {author} {\bibfnamefont {E.~N.}\ \bibnamefont {Fortson}},\ }\href
  {\doibase10.1103/PhysRevLett.57.3125} {\bibfield  {journal} {\bibinfo
  {journal} {Phys. Rev. Lett.}\ }\textbf {\bibinfo {volume} {57}},\ \bibinfo
  {pages} {3125} (\bibinfo {year} {1986})}\BibitemShut {NoStop}%
\bibitem [{\citenamefont {Bear}\ \textit {et~al.}(2000)\citenamefont {Bear},
  \citenamefont {Stoner}, \citenamefont {Walsworth}, \citenamefont
  {Kosteleck\'y},\ and\ \citenamefont {Lane}}]{Bear00}%
  \BibitemOpen
  \bibfield  {author} {\bibinfo {author} {\bibfnamefont {D.}~\bibnamefont
  {Bear}}, \bibinfo {author} {\bibfnamefont {R.~E.}\ \bibnamefont {Stoner}},
  \bibinfo {author} {\bibfnamefont {R.~L.}\ \bibnamefont {Walsworth}}, \bibinfo
  {author} {\bibfnamefont {V.~A.}\ \bibnamefont {Kosteleck\'y}}, \ and\
  \bibinfo {author} {\bibfnamefont {C.~D.}\ \bibnamefont {Lane}},\ }\href
  {\doibase10.1103/PhysRevLett.85.5038} {\bibfield  {journal} {\bibinfo
  {journal} {Phys. Rev. Lett.}\ }\textbf {\bibinfo {volume} {85}},\ \bibinfo
  {pages} {5038} (\bibinfo {year} {2000})}\BibitemShut {NoStop}%
\bibitem [{\citenamefont {Can\`e}\ \textit {et~al.}(2004)\citenamefont
  {Can\`e}, \citenamefont {Bear}, \citenamefont {Phillips}, \citenamefont
  {Rosen}, \citenamefont {Smallwood}, \citenamefont {Stoner}, \citenamefont
  {Walsworth},\ and\ \citenamefont {Kosteleck\'y}}]{Cane04}%
  \BibitemOpen
  \bibfield  {author} {\bibinfo {author} {\bibfnamefont {F.}~\bibnamefont
  {Can\`e}}, \bibinfo {author} {\bibfnamefont {D.}~\bibnamefont {Bear}},
  \bibinfo {author} {\bibfnamefont {D.~F.}\ \bibnamefont {Phillips}}, \bibinfo
  {author} {\bibfnamefont {M.~S.}\ \bibnamefont {Rosen}}, \bibinfo {author}
  {\bibfnamefont {C.~L.}\ \bibnamefont {Smallwood}}, \bibinfo {author}
  {\bibfnamefont {R.~E.}\ \bibnamefont {Stoner}}, \bibinfo {author}
  {\bibfnamefont {R.~L.}\ \bibnamefont {Walsworth}}, \ and\ \bibinfo {author}
  {\bibfnamefont {V.~A.}\ \bibnamefont {Kosteleck\'y}},\ }\href
  {\doibase10.1103/PhysRevLett.93.230801} {\bibfield  {journal} {\bibinfo
  {journal} {Phys. Rev. Lett.}\ }\textbf {\bibinfo {volume} {93}},\ \bibinfo
  {pages} {230801} (\bibinfo {year} {2004})}\BibitemShut {NoStop}%
\bibitem [{\citenamefont {Brown}\ \textit {et~al.}(2010)\citenamefont {Brown},
  \citenamefont {Smullin}, \citenamefont {Kornack},\ and\ \citenamefont
  {Romalis}}]{Brown10}%
  \BibitemOpen
  \bibfield  {author} {\bibinfo {author} {\bibfnamefont {J.~M.}\ \bibnamefont
  {Brown}}, \bibinfo {author} {\bibfnamefont {S.~J.}\ \bibnamefont {Smullin}},
  \bibinfo {author} {\bibfnamefont {T.~W.}\ \bibnamefont {Kornack}}, \ and\
  \bibinfo {author} {\bibfnamefont {M.~V.}\ \bibnamefont {Romalis}},\ }\href
  {\doibase10.1103/PhysRevLett.105.151604} {\bibfield  {journal} {\bibinfo
  {journal} {Phys. Rev. Lett.}\ }\textbf {\bibinfo {volume} {105}},\ \bibinfo
  {pages} {151604} (\bibinfo {year} {2010})}\BibitemShut {NoStop}%
\bibitem [{\citenamefont {Smiciklas}\ \textit {et~al.}(2011)\citenamefont
  {Smiciklas}, \citenamefont {Brown}, \citenamefont {Cheuk}, \citenamefont
  {Smullin},\ and\ \citenamefont {Romalis}}]{Smiciklas11}%
  \BibitemOpen
  \bibfield  {author} {\bibinfo {author} {\bibfnamefont {M.}~\bibnamefont
  {Smiciklas}}, \bibinfo {author} {\bibfnamefont {J.~M.}\ \bibnamefont
  {Brown}}, \bibinfo {author} {\bibfnamefont {L.~W.}\ \bibnamefont {Cheuk}},
  \bibinfo {author} {\bibfnamefont {S.~J.}\ \bibnamefont {Smullin}}, \ and\
  \bibinfo {author} {\bibfnamefont {M.~V.}\ \bibnamefont {Romalis}},\ }\href
  {\doibase10.1103/PhysRevLett.107.171604} {\bibfield  {journal} {\bibinfo
  {journal} {Phys. Rev. Lett.}\ }\textbf {\bibinfo {volume} {107}},\ \bibinfo
  {pages} {171604} (\bibinfo {year} {2011})}\BibitemShut {NoStop}%
\bibitem [{\citenamefont {Allmendinger}\ \textit {et~al.}(2014)\citenamefont
  {Allmendinger}, \citenamefont {Heil}, \citenamefont {Karpuk}, \citenamefont
  {Kilian}, \citenamefont {Scharth}, \citenamefont {Schmidt}, \citenamefont
  {Schnabel}, \citenamefont {Sobolev},\ and\ \citenamefont
  {Tullney}}]{Allmendinger14}%
  \BibitemOpen
  \bibfield  {author} {\bibinfo {author} {\bibfnamefont {F.}~\bibnamefont
  {Allmendinger}}, \bibinfo {author} {\bibfnamefont {W.}~\bibnamefont {Heil}},
  \bibinfo {author} {\bibfnamefont {S.}~\bibnamefont {Karpuk}}, \bibinfo
  {author} {\bibfnamefont {W.}~\bibnamefont {Kilian}}, \bibinfo {author}
  {\bibfnamefont {A.}~\bibnamefont {Scharth}}, \bibinfo {author} {\bibfnamefont
  {U.}~\bibnamefont {Schmidt}}, \bibinfo {author} {\bibfnamefont
  {A.}~\bibnamefont {Schnabel}}, \bibinfo {author} {\bibfnamefont
  {Y.}~\bibnamefont {Sobolev}}, \ and\ \bibinfo {author} {\bibfnamefont
  {K.}~\bibnamefont {Tullney}},\ }\href
  {\doibase10.1103/PhysRevLett.112.110801} {\bibfield  {journal} {\bibinfo
  {journal} {Phys. Rev. Lett.}\ }\textbf {\bibinfo {volume} {112}},\ \bibinfo
  {pages} {110801} (\bibinfo {year} {2014})}\BibitemShut {NoStop}%
\bibitem [{\citenamefont {Woodman}\ \textit {et~al.}(1987)\citenamefont
  {Woodman}, \citenamefont {Franks},\ and\ \citenamefont
  {Richards}}]{Woodman87}%
  \BibitemOpen
  \bibfield  {author} {\bibinfo {author} {\bibfnamefont {K.}~\bibnamefont
  {Woodman}}, \bibinfo {author} {\bibfnamefont {P.}~\bibnamefont {Franks}}, \
  and\ \bibinfo {author} {\bibfnamefont {M.}~\bibnamefont {Richards}},\
  }\href@noop {} {\bibfield  {journal} {\bibinfo  {journal} {The Journal of
  Navigation}\ }\textbf {\bibinfo {volume} {40}},\ \bibinfo {pages} {366}
  (\bibinfo {year} {1987})}\BibitemShut {NoStop}%
\bibitem [{\citenamefont {Kornack}\ \textit {et~al.}(2005)\citenamefont
  {Kornack}, \citenamefont {Ghosh},\ and\ \citenamefont {Romalis}}]{Kornack05}%
  \BibitemOpen
  \bibfield  {author} {\bibinfo {author} {\bibfnamefont {T.~W.}\ \bibnamefont
  {Kornack}}, \bibinfo {author} {\bibfnamefont {R.~K.}\ \bibnamefont {Ghosh}},
  \ and\ \bibinfo {author} {\bibfnamefont {M.~V.}\ \bibnamefont {Romalis}},\
  }\href {\doibase10.1103/PhysRevLett.95.230801} {\bibfield  {journal}
  {\bibinfo  {journal} {Phys. Rev. Lett.}\ }\textbf {\bibinfo {volume} {95}},\
  \bibinfo {pages} {230801} (\bibinfo {year} {2005})}\BibitemShut {NoStop}%
\bibitem [{\citenamefont {Limes}\ \textit {et~al.}(2018)\citenamefont {Limes},
  \citenamefont {Sheng},\ and\ \citenamefont {Romalis}}]{Limes18}%
  \BibitemOpen
  \bibfield  {author} {\bibinfo {author} {\bibfnamefont {M.~E.}\ \bibnamefont
  {Limes}}, \bibinfo {author} {\bibfnamefont {D.}~\bibnamefont {Sheng}}, \ and\
  \bibinfo {author} {\bibfnamefont {M.~V.}\ \bibnamefont {Romalis}},\ }\href
  {\doibase10.1103/PhysRevLett.120.033401} {\bibfield  {journal} {\bibinfo
  {journal} {Phys. Rev. Lett.}\ }\textbf {\bibinfo {volume} {120}},\ \bibinfo
  {pages} {033401} (\bibinfo {year} {2018})}\BibitemShut {NoStop}%
\bibitem [{\citenamefont {Jiang}\ \textit {et~al.}(2018)\citenamefont {Jiang},
  \citenamefont {Quan}, \citenamefont {Li}, \citenamefont {Fan}, \citenamefont
  {Liu}, \citenamefont {Qin}, \citenamefont {Wan},\ and\ \citenamefont
  {Fang}}]{Jiang18}%
  \BibitemOpen
  \bibfield  {author} {\bibinfo {author} {\bibfnamefont {L.}~\bibnamefont
  {Jiang}}, \bibinfo {author} {\bibfnamefont {W.}~\bibnamefont {Quan}},
  \bibinfo {author} {\bibfnamefont {R.}~\bibnamefont {Li}}, \bibinfo {author}
  {\bibfnamefont {W.}~\bibnamefont {Fan}}, \bibinfo {author} {\bibfnamefont
  {F.}~\bibnamefont {Liu}}, \bibinfo {author} {\bibfnamefont {J.}~\bibnamefont
  {Qin}}, \bibinfo {author} {\bibfnamefont {S.}~\bibnamefont {Wan}}, \ and\
  \bibinfo {author} {\bibfnamefont {J.}~\bibnamefont {Fang}},\ }\href@noop {}
  {\bibfield  {journal} {\bibinfo  {journal} {Applied Physics Letters}\
  }\textbf {\bibinfo {volume} {112}},\ \bibinfo {pages} {054103} (\bibinfo
  {year} {2018})}\BibitemShut {NoStop}%
\bibitem [{\citenamefont {Kornack}\ and\ \citenamefont
  {Romalis}(2002)}]{Kornack02}%
  \BibitemOpen
  \bibfield  {author} {\bibinfo {author} {\bibfnamefont {T.~W.}\ \bibnamefont
  {Kornack}}\ and\ \bibinfo {author} {\bibfnamefont {M.~V.}\ \bibnamefont
  {Romalis}},\ }\href {\doibase10.1103/PhysRevLett.89.253002} {\bibfield
  {journal} {\bibinfo  {journal} {Phys. Rev. Lett.}\ }\textbf {\bibinfo
  {volume} {89}},\ \bibinfo {pages} {253002} (\bibinfo {year}
  {2002})}\BibitemShut {NoStop}%
\bibitem [{\citenamefont {Sheng}\ \textit {et~al.}(2014)\citenamefont {Sheng},
  \citenamefont {Kabcenell},\ and\ \citenamefont {Romalis}}]{Sheng14}%
  \BibitemOpen
  \bibfield  {author} {\bibinfo {author} {\bibfnamefont {D.}~\bibnamefont
  {Sheng}}, \bibinfo {author} {\bibfnamefont {A.}~\bibnamefont {Kabcenell}}, \
  and\ \bibinfo {author} {\bibfnamefont {M.~V.}\ \bibnamefont {Romalis}},\
  }\href {\doibase10.1103/PhysRevLett.113.163002} {\bibfield  {journal}
  {\bibinfo  {journal} {Phys. Rev. Lett.}\ }\textbf {\bibinfo {volume} {113}},\
  \bibinfo {pages} {163002} (\bibinfo {year} {2014})}\BibitemShut {NoStop}%
\bibitem [{\citenamefont {Ledbetter}\ \textit {et~al.}(2012)\citenamefont
  {Ledbetter}, \citenamefont {Pustelny}, \citenamefont {Budker}, \citenamefont
  {Romalis}, \citenamefont {Blanchard},\ and\ \citenamefont
  {Pines}}]{Ledbetter12}%
  \BibitemOpen
  \bibfield  {author} {\bibinfo {author} {\bibfnamefont {M.~P.}\ \bibnamefont
  {Ledbetter}}, \bibinfo {author} {\bibfnamefont {S.}~\bibnamefont {Pustelny}},
  \bibinfo {author} {\bibfnamefont {D.}~\bibnamefont {Budker}}, \bibinfo
  {author} {\bibfnamefont {M.~V.}\ \bibnamefont {Romalis}}, \bibinfo {author}
  {\bibfnamefont {J.~W.}\ \bibnamefont {Blanchard}}, \ and\ \bibinfo {author}
  {\bibfnamefont {A.}~\bibnamefont {Pines}},\ }\href
  {\doibase10.1103/PhysRevLett.108.243001} {\bibfield  {journal} {\bibinfo
  {journal} {Phys. Rev. Lett.}\ }\textbf {\bibinfo {volume} {108}},\ \bibinfo
  {pages} {243001} (\bibinfo {year} {2012})}\BibitemShut {NoStop}%
\bibitem [{\citenamefont {Wu}\ \textit {et~al.}(2018)\citenamefont {Wu},
  \citenamefont {Blanchard}, \citenamefont {Jackson~Kimball}, \citenamefont
  {Jiang},\ and\ \citenamefont {Budker}}]{Wu18}%
  \BibitemOpen
  \bibfield  {author} {\bibinfo {author} {\bibfnamefont {T.}~\bibnamefont
  {Wu}}, \bibinfo {author} {\bibfnamefont {J.~W.}\ \bibnamefont {Blanchard}},
  \bibinfo {author} {\bibfnamefont {D.~F.}\ \bibnamefont {Jackson~Kimball}},
  \bibinfo {author} {\bibfnamefont {M.}~\bibnamefont {Jiang}}, \ and\ \bibinfo
  {author} {\bibfnamefont {D.}~\bibnamefont {Budker}},\ }\href
  {\doibase10.1103/PhysRevLett.121.023202} {\bibfield  {journal} {\bibinfo
  {journal} {Phys. Rev. Lett.}\ }\textbf {\bibinfo {volume} {121}},\ \bibinfo
  {pages} {023202} (\bibinfo {year} {2018})}\BibitemShut {NoStop}%
\bibitem [{\citenamefont {Palacios}\ \textit {et~al.}(2018)\citenamefont
  {Palacios}, \citenamefont {Coop}, \citenamefont {Gomez}, \citenamefont
  {Vanderbruggen}, \citenamefont {de~Escobar}, \citenamefont {Jasperse},\ and\
  \citenamefont {Mitchell}}]{Palacios18}%
  \BibitemOpen
  \bibfield  {author} {\bibinfo {author} {\bibfnamefont {S.}~\bibnamefont
  {Palacios}}, \bibinfo {author} {\bibfnamefont {S.}~\bibnamefont {Coop}},
  \bibinfo {author} {\bibfnamefont {P.}~\bibnamefont {Gomez}}, \bibinfo
  {author} {\bibfnamefont {T.}~\bibnamefont {Vanderbruggen}}, \bibinfo {author}
  {\bibfnamefont {Y.~N.~M.}\ \bibnamefont {de~Escobar}}, \bibinfo {author}
  {\bibfnamefont {M.}~\bibnamefont {Jasperse}}, \ and\ \bibinfo {author}
  {\bibfnamefont {M.~W.}\ \bibnamefont {Mitchell}},\ }\href
  {https://iopscience.iop.org/article/10.1088/1367-2630/aab2a0} {\bibfield
  {journal} {\bibinfo  {journal} {New Journal of Physics}\ }\textbf {\bibinfo
  {volume} {20}},\ \bibinfo {pages} {053008} (\bibinfo {year}
  {2018})}\BibitemShut {NoStop}%
\bibitem [{\citenamefont {Vengalattore}\ \textit {et~al.}(2007)\citenamefont
  {Vengalattore}, \citenamefont {Higbie}, \citenamefont {Leslie}, \citenamefont
  {Guzman}, \citenamefont {Sadler},\ and\ \citenamefont
  {Stamper-Kurn}}]{VengalattorePRL2007}%
  \BibitemOpen
  \bibfield  {author} {\bibinfo {author} {\bibfnamefont {M.}~\bibnamefont
  {Vengalattore}}, \bibinfo {author} {\bibfnamefont {J.~M.}\ \bibnamefont
  {Higbie}}, \bibinfo {author} {\bibfnamefont {S.~R.}\ \bibnamefont {Leslie}},
  \bibinfo {author} {\bibfnamefont {J.}~\bibnamefont {Guzman}}, \bibinfo
  {author} {\bibfnamefont {L.~E.}\ \bibnamefont {Sadler}}, \ and\ \bibinfo
  {author} {\bibfnamefont {D.~M.}\ \bibnamefont {Stamper-Kurn}},\ }\href
  {\doibase10.1103/PhysRevLett.98.200801} {\bibfield  {journal} {\bibinfo
  {journal} {Phys. Rev. Lett.}\ }\textbf {\bibinfo {volume} {98}},\ \bibinfo
  {pages} {200801} (\bibinfo {year} {2007})}\BibitemShut {NoStop}%
\bibitem [{\citenamefont {Vanderbruggen}\ \textit {et~al.}(2015)\citenamefont
  {Vanderbruggen}, \citenamefont {{\'A}lvarez}, \citenamefont {Coop},
  \citenamefont {de~Escobar},\ and\ \citenamefont
  {Mitchell}}]{VanderbruggenEPL2015}%
  \BibitemOpen
  \bibfield  {author} {\bibinfo {author} {\bibfnamefont {T.}~\bibnamefont
  {Vanderbruggen}}, \bibinfo {author} {\bibfnamefont {S.~P.}\ \bibnamefont
  {{\'A}lvarez}}, \bibinfo {author} {\bibfnamefont {S.}~\bibnamefont {Coop}},
  \bibinfo {author} {\bibfnamefont {N.~M.}\ \bibnamefont {de~Escobar}}, \ and\
  \bibinfo {author} {\bibfnamefont {M.~W.}\ \bibnamefont {Mitchell}},\ }\href
  {http://stacks.iop.org/0295-5075/111/i=6/a=66001} {\bibfield  {journal}
  {\bibinfo  {journal} {Europhys Lett}\ }\textbf {\bibinfo {volume} {111}},\
  \bibinfo {pages} {66001} (\bibinfo {year} {2015})}\BibitemShut {NoStop}%
\bibitem [{\citenamefont {Gomez}\ \textit {et~al.}(2019)\citenamefont {Gomez},
  \citenamefont {Mazzinghi}, \citenamefont {Martin}, \citenamefont {Coop},
  \citenamefont {Palacios},\ and\ \citenamefont {Mitchell}}]{Gomez19}%
  \BibitemOpen
  \bibfield  {author} {\bibinfo {author} {\bibfnamefont {P.}~\bibnamefont
  {Gomez}}, \bibinfo {author} {\bibfnamefont {C.}~\bibnamefont {Mazzinghi}},
  \bibinfo {author} {\bibfnamefont {F.}~\bibnamefont {Martin}}, \bibinfo
  {author} {\bibfnamefont {S.}~\bibnamefont {Coop}}, \bibinfo {author}
  {\bibfnamefont {S.}~\bibnamefont {Palacios}}, \ and\ \bibinfo {author}
  {\bibfnamefont {M.~W.}\ \bibnamefont {Mitchell}},\ }\href
  {\doibase10.1103/PhysRevA.100.032704} {\bibfield  {journal} {\bibinfo
  {journal} {Phys. Rev. A}\ }\textbf {\bibinfo {volume} {100}},\ \bibinfo
  {pages} {032704} (\bibinfo {year} {2019})}\BibitemShut {NoStop}%
\bibitem [{\citenamefont {Schmaljohann}\ \textit {et~al.}(2004)\citenamefont
  {Schmaljohann}, \citenamefont {Erhard}, \citenamefont {Kronj\"ager},
  \citenamefont {Kottke}, \citenamefont {van Staa}, \citenamefont
  {Cacciapuoti}, \citenamefont {Arlt}, \citenamefont {Bongs},\ and\
  \citenamefont {Sengstock}}]{Schmaljohann04}%
  \BibitemOpen
  \bibfield  {author} {\bibinfo {author} {\bibfnamefont {H.}~\bibnamefont
  {Schmaljohann}}, \bibinfo {author} {\bibfnamefont {M.}~\bibnamefont
  {Erhard}}, \bibinfo {author} {\bibfnamefont {J.}~\bibnamefont {Kronj\"ager}},
  \bibinfo {author} {\bibfnamefont {M.}~\bibnamefont {Kottke}}, \bibinfo
  {author} {\bibfnamefont {S.}~\bibnamefont {van Staa}}, \bibinfo {author}
  {\bibfnamefont {L.}~\bibnamefont {Cacciapuoti}}, \bibinfo {author}
  {\bibfnamefont {J.~J.}\ \bibnamefont {Arlt}}, \bibinfo {author}
  {\bibfnamefont {K.}~\bibnamefont {Bongs}}, \ and\ \bibinfo {author}
  {\bibfnamefont {K.}~\bibnamefont {Sengstock}},\ }\href
  {\doibase10.1103/PhysRevLett.92.040402} {\bibfield  {journal} {\bibinfo
  {journal} {Phys. Rev. Lett.}\ }\textbf {\bibinfo {volume} {92}},\ \bibinfo
  {pages} {040402} (\bibinfo {year} {2004})}\BibitemShut {NoStop}%
\bibitem [{\citenamefont {Tojo}\ \textit {et~al.}(2009)\citenamefont {Tojo},
  \citenamefont {Hayashi}, \citenamefont {Tanabe}, \citenamefont {Hirano},
  \citenamefont {Kawaguchi}, \citenamefont {Saito},\ and\ \citenamefont
  {Ueda}}]{Tojo09}%
  \BibitemOpen
  \bibfield  {author} {\bibinfo {author} {\bibfnamefont {S.}~\bibnamefont
  {Tojo}}, \bibinfo {author} {\bibfnamefont {T.}~\bibnamefont {Hayashi}},
  \bibinfo {author} {\bibfnamefont {T.}~\bibnamefont {Tanabe}}, \bibinfo
  {author} {\bibfnamefont {T.}~\bibnamefont {Hirano}}, \bibinfo {author}
  {\bibfnamefont {Y.}~\bibnamefont {Kawaguchi}}, \bibinfo {author}
  {\bibfnamefont {H.}~\bibnamefont {Saito}}, \ and\ \bibinfo {author}
  {\bibfnamefont {M.}~\bibnamefont {Ueda}},\ }\href
  {\doibase10.1103/PhysRevA.80.042704} {\bibfield  {journal} {\bibinfo
  {journal} {Phys. Rev. A}\ }\textbf {\bibinfo {volume} {80}},\ \bibinfo
  {pages} {042704} (\bibinfo {year} {2009})}\BibitemShut {NoStop}%
\bibitem [{\citenamefont {Pu}\ \textit {et~al.}(1999)\citenamefont {Pu},
  \citenamefont {Law}, \citenamefont {Raghavan}, \citenamefont {Eberly},\ and\
  \citenamefont {Bigelow}}]{Pu99}%
  \BibitemOpen
  \bibfield  {author} {\bibinfo {author} {\bibfnamefont {H.}~\bibnamefont
  {Pu}}, \bibinfo {author} {\bibfnamefont {C.~K.}\ \bibnamefont {Law}},
  \bibinfo {author} {\bibfnamefont {S.}~\bibnamefont {Raghavan}}, \bibinfo
  {author} {\bibfnamefont {J.~H.}\ \bibnamefont {Eberly}}, \ and\ \bibinfo
  {author} {\bibfnamefont {N.~P.}\ \bibnamefont {Bigelow}},\ }\href
  {\doibase10.1103/PhysRevA.60.1463} {\bibfield  {journal} {\bibinfo  {journal}
  {Phys. Rev. A}\ }\textbf {\bibinfo {volume} {60}},\ \bibinfo {pages} {1463}
  (\bibinfo {year} {1999})}\BibitemShut {NoStop}%
\bibitem [{\citenamefont {Kawaguchi}\ and\ \citenamefont
  {Ueda}(2012)}]{Kawaguchi12}%
  \BibitemOpen
  \bibfield  {author} {\bibinfo {author} {\bibfnamefont {Y.}~\bibnamefont
  {Kawaguchi}}\ and\ \bibinfo {author} {\bibfnamefont {M.}~\bibnamefont
  {Ueda}},\ }\href
  {http://www.sciencedirect.com/science/article/pii/S0370157312002098}
  {\bibfield  {journal} {\bibinfo  {journal} {Physics Reports}\ }\textbf
  {\bibinfo {volume} {520}},\ \bibinfo {pages} {253} (\bibinfo {year}
  {2012})}\BibitemShut {NoStop}%
\bibitem [{\citenamefont {Koschorreck}\ \textit {et~al.}(2010)\citenamefont
  {Koschorreck}, \citenamefont {Napolitano}, \citenamefont {Dubost},\ and\
  \citenamefont {Mitchell}}]{Koschorreck10}%
  \BibitemOpen
  \bibfield  {author} {\bibinfo {author} {\bibfnamefont {M.}~\bibnamefont
  {Koschorreck}}, \bibinfo {author} {\bibfnamefont {M.}~\bibnamefont
  {Napolitano}}, \bibinfo {author} {\bibfnamefont {B.}~\bibnamefont {Dubost}},
  \ and\ \bibinfo {author} {\bibfnamefont {M.~W.}\ \bibnamefont {Mitchell}},\
  }\href {\doibase10.1103/PhysRevLett.105.093602} {\bibfield  {journal}
  {\bibinfo  {journal} {Phys. Rev. Lett.}\ }\textbf {\bibinfo {volume} {105}},\
  \bibinfo {pages} {093602} (\bibinfo {year} {2010})}\BibitemShut {NoStop}%
\bibitem [{\citenamefont {Geremia}\ \textit {et~al.}(2006)\citenamefont
  {Geremia}, \citenamefont {Stockton},\ and\ \citenamefont
  {Mabuchi}}]{Geremia06}%
  \BibitemOpen
  \bibfield  {author} {\bibinfo {author} {\bibfnamefont {J.~M.}\ \bibnamefont
  {Geremia}}, \bibinfo {author} {\bibfnamefont {J.~K.}\ \bibnamefont
  {Stockton}}, \ and\ \bibinfo {author} {\bibfnamefont {H.}~\bibnamefont
  {Mabuchi}},\ }\href {\doibase10.1103/PhysRevA.73.042112} {\bibfield
  {journal} {\bibinfo  {journal} {Phys. Rev. A}\ }\textbf {\bibinfo {volume}
  {73}},\ \bibinfo {pages} {042112} (\bibinfo {year} {2006})}\BibitemShut
  {NoStop}%
\bibitem [{\citenamefont {Ciurana}\ \textit {et~al.}(2016)\citenamefont
  {Ciurana}, \citenamefont {Colangelo}, \citenamefont {Sewell},\ and\
  \citenamefont {Mitchell}}]{Martin16}%
  \BibitemOpen
  \bibfield  {author} {\bibinfo {author} {\bibfnamefont {F.~M.}\ \bibnamefont
  {Ciurana}}, \bibinfo {author} {\bibfnamefont {G.}~\bibnamefont {Colangelo}},
  \bibinfo {author} {\bibfnamefont {R.~J.}\ \bibnamefont {Sewell}}, \ and\
  \bibinfo {author} {\bibfnamefont {M.~W.}\ \bibnamefont {Mitchell}},\ }\href
  {\doibase10.1364/OL.41.002946} {\bibfield  {journal} {\bibinfo  {journal}
  {Opt. Lett.}\ }\textbf {\bibinfo {volume} {41}},\ \bibinfo {pages} {2946}
  (\bibinfo {year} {2016})}\BibitemShut {NoStop}%
\bibitem [{\citenamefont {Irikura}\ \textit {et~al.}(2018)\citenamefont
  {Irikura}, \citenamefont {Eto}, \citenamefont {Hirano},\ and\ \citenamefont
  {Saito}}]{Saito18}%
  \BibitemOpen
  \bibfield  {author} {\bibinfo {author} {\bibfnamefont {N.}~\bibnamefont
  {Irikura}}, \bibinfo {author} {\bibfnamefont {Y.}~\bibnamefont {Eto}},
  \bibinfo {author} {\bibfnamefont {T.}~\bibnamefont {Hirano}}, \ and\ \bibinfo
  {author} {\bibfnamefont {H.}~\bibnamefont {Saito}},\ }\href
  {\doibase10.1103/PhysRevA.97.023622} {\bibfield  {journal} {\bibinfo
  {journal} {Phys. Rev. A}\ }\textbf {\bibinfo {volume} {97}},\ \bibinfo
  {pages} {023622} (\bibinfo {year} {2018})}\BibitemShut {NoStop}%
\bibitem [{\citenamefont {Berry}\ \textit {et~al.}(2001)\citenamefont {Berry},
  \citenamefont {Wiseman},\ and\ \citenamefont {Breslin}}]{Berry01}%
  \BibitemOpen
  \bibfield  {author} {\bibinfo {author} {\bibfnamefont {D.~W.}\ \bibnamefont
  {Berry}}, \bibinfo {author} {\bibfnamefont {H.~M.}\ \bibnamefont {Wiseman}},
  \ and\ \bibinfo {author} {\bibfnamefont {J.~K.}\ \bibnamefont {Breslin}},\
  }\href {\doibase10.1103/PhysRevA.63.053804} {\bibfield  {journal} {\bibinfo
  {journal} {Phys. Rev. A}\ }\textbf {\bibinfo {volume} {63}},\ \bibinfo
  {pages} {053804} (\bibinfo {year} {2001})}\BibitemShut {NoStop}%
\bibitem [{\citenamefont {Lodewyck}\ \textit {et~al.}(2009)\citenamefont
  {Lodewyck}, \citenamefont {Westergaard},\ and\ \citenamefont
  {Lemonde}}]{LodewyckPRA2009}%
  \BibitemOpen
  \bibfield  {author} {\bibinfo {author} {\bibfnamefont {J.}~\bibnamefont
  {Lodewyck}}, \bibinfo {author} {\bibfnamefont {P.~G.}\ \bibnamefont
  {Westergaard}}, \ and\ \bibinfo {author} {\bibfnamefont {P.}~\bibnamefont
  {Lemonde}},\ }\href {\doibase10.1103/PhysRevA.79.061401} {\bibfield
  {journal} {\bibinfo  {journal} {Phys. Rev. A}\ }\textbf {\bibinfo {volume}
  {79}},\ \bibinfo {pages} {061401} (\bibinfo {year} {2009})}\BibitemShut
  {NoStop}%
\bibitem [{\citenamefont {Schleier-Smith}\ \textit {et~al.}(2010)\citenamefont
  {Schleier-Smith}, \citenamefont {Leroux},\ and\ \citenamefont
  {Vuletic}}]{Schleier-SmithPRL2010}%
  \BibitemOpen
  \bibfield  {author} {\bibinfo {author} {\bibfnamefont {M.~H.}\ \bibnamefont
  {Schleier-Smith}}, \bibinfo {author} {\bibfnamefont {I.~D.}\ \bibnamefont
  {Leroux}}, \ and\ \bibinfo {author} {\bibfnamefont {V.}~\bibnamefont
  {Vuletic}},\ }\href {\doibase10.1103/PhysRevLett.104.073604} {\bibfield
  {journal} {\bibinfo  {journal} {Phys. Rev. Lett.}\ }\textbf {\bibinfo
  {volume} {104}},\ \bibinfo {pages} {073604} (\bibinfo {year}
  {2010})}\BibitemShut {NoStop}%
\bibitem [{\citenamefont {Lee}\ \textit {et~al.}(2014)\citenamefont {Lee},
  \citenamefont {Vrijsen}, \citenamefont {Teper}, \citenamefont {Hosten},\ and\
  \citenamefont {Kasevich}}]{LeeOL2014}%
  \BibitemOpen
  \bibfield  {author} {\bibinfo {author} {\bibfnamefont {J.}~\bibnamefont
  {Lee}}, \bibinfo {author} {\bibfnamefont {G.}~\bibnamefont {Vrijsen}},
  \bibinfo {author} {\bibfnamefont {I.}~\bibnamefont {Teper}}, \bibinfo
  {author} {\bibfnamefont {O.}~\bibnamefont {Hosten}}, \ and\ \bibinfo {author}
  {\bibfnamefont {M.~A.}\ \bibnamefont {Kasevich}},\ }\href
  {\doibase10.1364/OL.39.004005} {\bibfield  {journal} {\bibinfo  {journal}
  {Opt. Lett.}\ }\textbf {\bibinfo {volume} {39}},\ \bibinfo {pages} {4005}
  (\bibinfo {year} {2014})}\BibitemShut {NoStop}%
\bibitem [{\citenamefont {Raffelt}(2012)}]{RaffeltPRD2012}%
  \BibitemOpen
  \bibfield  {author} {\bibinfo {author} {\bibfnamefont {G.}~\bibnamefont
  {Raffelt}},\ }\href {\doibase10.1103/PhysRevD.86.015001} {\bibfield
  {journal} {\bibinfo  {journal} {Phys. Rev. D}\ }\textbf {\bibinfo {volume}
  {86}},\ \bibinfo {pages} {015001} (\bibinfo {year} {2012})}\BibitemShut
  {NoStop}%
\bibitem [{\citenamefont {Petukhov}\ \textit {et~al.}(2010)\citenamefont
  {Petukhov}, \citenamefont {Pignol}, \citenamefont {Jullien},\ and\
  \citenamefont {Andersen}}]{PetukhovPRL2010}%
  \BibitemOpen
  \bibfield  {author} {\bibinfo {author} {\bibfnamefont {A.~K.}\ \bibnamefont
  {Petukhov}}, \bibinfo {author} {\bibfnamefont {G.}~\bibnamefont {Pignol}},
  \bibinfo {author} {\bibfnamefont {D.}~\bibnamefont {Jullien}}, \ and\
  \bibinfo {author} {\bibfnamefont {K.~H.}\ \bibnamefont {Andersen}},\ }\href
  {\doibase10.1103/PhysRevLett.105.170401} {\bibfield  {journal} {\bibinfo
  {journal} {Phys. Rev. Lett.}\ }\textbf {\bibinfo {volume} {105}},\ \bibinfo
  {pages} {170401} (\bibinfo {year} {2010})}\BibitemShut {NoStop}%
\bibitem [{\citenamefont {Serebrov}\ \textit {et~al.}(2010)\citenamefont
  {Serebrov}, \citenamefont {Zimmer}, \citenamefont {Geltenbort}, \citenamefont
  {Fomin}, \citenamefont {Ivanov}, \citenamefont {Kolomensky}, \citenamefont
  {Krasnoshekova}, \citenamefont {Lasakov}, \citenamefont {Lobashev},
  \citenamefont {Pirozhkov}, \citenamefont {Varlamov}, \citenamefont
  {Vasiliev}, \citenamefont {Zherebtsov}, \citenamefont {Aleksandrov},
  \citenamefont {Dmitriev},\ and\ \citenamefont {Dovator}}]{SerebrovJETPL2010}%
  \BibitemOpen
  \bibfield  {author} {\bibinfo {author} {\bibfnamefont {A.~P.}\ \bibnamefont
  {Serebrov}}, \bibinfo {author} {\bibfnamefont {O.}~\bibnamefont {Zimmer}},
  \bibinfo {author} {\bibfnamefont {P.}~\bibnamefont {Geltenbort}}, \bibinfo
  {author} {\bibfnamefont {A.~K.}\ \bibnamefont {Fomin}}, \bibinfo {author}
  {\bibfnamefont {S.~N.}\ \bibnamefont {Ivanov}}, \bibinfo {author}
  {\bibfnamefont {E.~A.}\ \bibnamefont {Kolomensky}}, \bibinfo {author}
  {\bibfnamefont {I.~A.}\ \bibnamefont {Krasnoshekova}}, \bibinfo {author}
  {\bibfnamefont {M.~S.}\ \bibnamefont {Lasakov}}, \bibinfo {author}
  {\bibfnamefont {V.~M.}\ \bibnamefont {Lobashev}}, \bibinfo {author}
  {\bibfnamefont {A.~N.}\ \bibnamefont {Pirozhkov}}, \bibinfo {author}
  {\bibfnamefont {V.~E.}\ \bibnamefont {Varlamov}}, \bibinfo {author}
  {\bibfnamefont {A.~V.}\ \bibnamefont {Vasiliev}}, \bibinfo {author}
  {\bibfnamefont {O.~M.}\ \bibnamefont {Zherebtsov}}, \bibinfo {author}
  {\bibfnamefont {E.~B.}\ \bibnamefont {Aleksandrov}}, \bibinfo {author}
  {\bibfnamefont {S.~P.}\ \bibnamefont {Dmitriev}}, \ and\ \bibinfo {author}
  {\bibfnamefont {N.~A.}\ \bibnamefont {Dovator}},\ }\href
  {\doibase10.1134/S0021364010010029} {\bibfield  {journal} {\bibinfo
  {journal} {JETP Letters}\ }\textbf {\bibinfo {volume} {91}},\ \bibinfo
  {pages} {6} (\bibinfo {year} {2010})}\BibitemShut {NoStop}%
\bibitem [{\citenamefont {Leslie}\ \textit {et~al.}(2009)\citenamefont
  {Leslie}, \citenamefont {Guzman}, \citenamefont {Vengalattore}, \citenamefont
  {Sau}, \citenamefont {Cohen},\ and\ \citenamefont {Stamper-Kurn}}]{Leslie09}%
  \BibitemOpen
  \bibfield  {author} {\bibinfo {author} {\bibfnamefont {S.~R.}\ \bibnamefont
  {Leslie}}, \bibinfo {author} {\bibfnamefont {J.}~\bibnamefont {Guzman}},
  \bibinfo {author} {\bibfnamefont {M.}~\bibnamefont {Vengalattore}}, \bibinfo
  {author} {\bibfnamefont {J.~D.}\ \bibnamefont {Sau}}, \bibinfo {author}
  {\bibfnamefont {M.~L.}\ \bibnamefont {Cohen}}, \ and\ \bibinfo {author}
  {\bibfnamefont {D.~M.}\ \bibnamefont {Stamper-Kurn}},\ }\href
  {\doibase10.1103/PhysRevA.79.043631} {\bibfield  {journal} {\bibinfo
  {journal} {Phys. Rev. A}\ }\textbf {\bibinfo {volume} {79}},\ \bibinfo
  {pages} {043631} (\bibinfo {year} {2009})}\BibitemShut {NoStop}%
\bibitem [{\citenamefont {Klempt}\ \textit {et~al.}(2010)\citenamefont
  {Klempt}, \citenamefont {Topic}, \citenamefont {Gebreyesus}, \citenamefont
  {Scherer}, \citenamefont {Henninger}, \citenamefont {Hyllus}, \citenamefont
  {Ertmer}, \citenamefont {Santos},\ and\ \citenamefont {Arlt}}]{Klempt10}%
  \BibitemOpen
  \bibfield  {author} {\bibinfo {author} {\bibfnamefont {C.}~\bibnamefont
  {Klempt}}, \bibinfo {author} {\bibfnamefont {O.}~\bibnamefont {Topic}},
  \bibinfo {author} {\bibfnamefont {G.}~\bibnamefont {Gebreyesus}}, \bibinfo
  {author} {\bibfnamefont {M.}~\bibnamefont {Scherer}}, \bibinfo {author}
  {\bibfnamefont {T.}~\bibnamefont {Henninger}}, \bibinfo {author}
  {\bibfnamefont {P.}~\bibnamefont {Hyllus}}, \bibinfo {author} {\bibfnamefont
  {W.}~\bibnamefont {Ertmer}}, \bibinfo {author} {\bibfnamefont
  {L.}~\bibnamefont {Santos}}, \ and\ \bibinfo {author} {\bibfnamefont {J.~J.}\
  \bibnamefont {Arlt}},\ }\href {\doibase10.1103/PhysRevLett.104.195303}
  {\bibfield  {journal} {\bibinfo  {journal} {Phys. Rev. Lett.}\ }\textbf
  {\bibinfo {volume} {104}},\ \bibinfo {pages} {195303} (\bibinfo {year}
  {2010})}\BibitemShut {NoStop}%
\end{thebibliography}%

\end{document}